\documentstyle[preprint,revtex]{aps}	
\begin{document}
%
\renewcommand{\thepage}{\roman{page}}
\setcounter{page}{1}

\preprint{FSU-HEP-930801}
\preprint{LBL-34420}
\preprint{August 1, 1993}
\begin{title}
Impact of QCD Corrections on the \\ Search for the
Intermediate Mass Higgs Boson$\;$\footnote[1]{{\em
This work was supported by the Director, Office of Energy
Research, Office of High Energy and Nuclear Physics, Division of High
Energy Physics of the U.S. Department of Energy under Contract
DE-AC03-76SF00098.}}
\end{title}
\author{B.~Bailey}
\begin{instit}
Department of Physics, B-159,
Florida State University\\
Tallahassee, Florida 32306
\end{instit}
\author{D. Graudenz}
\begin{instit}
Theoretical Physics Group,
Lawrence Berkeley Laboratory, University of California\\
1 Cyclotron Road, Berkeley, California 94720
\end{instit}
\begin{abstract}
Using next-to-leading-log calculations of Higgs production and hadronic two
photon production,
a signal and background study in the intermediate mass
range of the Higgs boson is
done for $\sqrt{s}=$ 40 and 14 TeV. The effects of realistic
cuts, including photon
isolation, are examined.
\end{abstract}
\newpage
\renewcommand{\thepage}{\roman{page}}
\setcounter{page}{2}
\mbox{ }

\vskip 1in

\begin{center}
{\bf Disclaimer}
\end{center}

\vskip .2in

\begin{scriptsize}
\begin{quotation}
This document was prepared as an account of work sponsored by the United
States Government.  Neither the United States Government nor any agency
thereof, nor The Regents of the University of California, nor any of their
employees, makes any warranty, express or implied, or assumes any legal
liability or responsibility for the accuracy, completeness, or usefulness
of any information, apparatus, product, or process disclosed, or represents
that its use would not infringe privately owned rights.  Reference herein
to any specific commercial products process, or service by its trade name,
trademark, manufacturer, or otherwise, does not necessarily constitute or
imply its endorsement, recommendation, or favoring by the United States
Government or any agency thereof, or The Regents of the University of
California.  The views and opinions of authors expressed herein do not
necessarily state or reflect those of the United States Government or any
agency thereof of The Regents of the University of California and shall
not be used for advertising or product endorsement purposes.
\end{quotation}
\end{scriptsize}

\vskip 2in

\begin{center}
\begin{small}
{\it Lawrence Berkeley Laboratory is an equal opportunity employer.}
\end{small}
\end{center}

\newpage
\renewcommand{\thepage}{\arabic{page}}

\setcounter{page}{1}

%
%
\begin{narrowtext}
\section{Introduction}

The Higgs boson and the top-quark are the remaining missing pieces of the
Standard Model.
The top-quark, if its mass is less than 250 GeV, is
expected to be discovered at
the Tevatron~\cite{TOP} and
LEP has set a lower bound on the Higgs mass of $m_{H}>$ 57 GeV~\cite{LEP}.
LEP II, with a center
of mass energy of 180 Gev, will be able to extend the
search to 90 Gev~\cite{LEPT}.
If $m_{H}>$ 90 GeV, high energy colliders such as the SSC and/or LHC will
be required to extend the search.

The dominant production mechanism at hadron supercolliders is expected to be
$gg\rightarrow H$~\cite{HHG}.
The dominant decay mode depends on the Higgs mass. If $m_{H}$ is
greater than twice the $Z$ mass
the four lepton decay mode, $H\rightarrow ZZ \rightarrow\;$4$l$, should
be observable~\cite{HHG}.
If the Higgs mass lies in the ``intermediate mass region'',
80 GeV $< m_{H}<$ 2$M_{Z}$, QCD
backgrounds overwhelm the main decay mode $H\rightarrow b \bar{b}$
and the rare processes
$H\rightarrow ZZ^{*}$ and $H\rightarrow \gamma\gamma$ become the decay
modes of choice~\cite{GKW}.
The decay $H\rightarrow ZZ^{*}$ occurs at observable
rates for $m_{H}>$ 130 GeV and
$H\rightarrow \gamma\gamma$ occurs at observable rates for the
entire intermediate mass region.
Unfortunately, the two photon decay mode is plagued by a large
background~\cite{GGJH} and
detection of the inclusive process $pp\to H\rightarrow\gamma\gamma$
will require detectors
with excellent $\gamma\gamma$ mass resolution~\cite{GEMTDR}.

Alternative production mechanisms which eliminate the large two
photon background by the inclusion
of a final state lepton have been studied~\cite{KKS,TTH,HANW,HBBO}. These
production mechanisms include
associated $W$ $H$, or $t \bar{t}$ $H$ production with
$H\rightarrow \gamma\gamma$.
Including cuts, the expected number of such events per year at the
SSC (LHC), assuming $\sqrt{s}=$ 40 (14) TeV and a luminosity
of 10 $fb^{-1}$, is $\sim$ 20 (4).
By comparison, for $m_{H}=$ 140 GeV the expected number of
$pp\rightarrow H\rightarrow \gamma\gamma$ events per year is
$\sim$ 700 (300) at the SSC (LHC). The Higgs can be discovered via the
$l\gamma\gamma$ signal but confirmation
of the discovery in the $\gamma\gamma$ channel would provide
a margin of certainty.
Clearly the $\gamma\gamma$ channel requires precise knowledge of the two
photon signal and background.

Recently, next-to-leading-log (NLL) calculations of Higgs
production~\cite{HQCD1,HQCD2,HQCD3}
and photon pair production have been performed~\cite{GG,GG2}.
The photon pair production calculation was performed in a
Monte Carlo environment.
The flexibility of the Monte Carlo calculation allows a thorough study of
the effect of various kinematic
and isolation cuts on the two photon background. In order
to make useful comparisons between signal and
background, the signal $pp\rightarrow H\rightarrow \gamma\gamma$ was
recalculated using the
NLL Monte Carlo formalism. In this paper the effect
of kinematic and isolation
cuts on the signal and background is examined at the NLL level
(for the intermediate mass region).
Results are presented for both the SSC and LHC.

The rest of this paper is organized as follows. In section II, details
of the calculation as well as input
parameters are discussed. In section III, numerical results are presented.
In section IV, a brief summary
and conclusions are presented.

\section{Calculation}
The Monte Carlo formalism for next-to-leading-log (NLL) calculations has
been described in
detail in Refs.~\cite{GG,GG2,NLLMC}. The explicit details for the two photon
calculation can be found in Ref.~\cite{GG} and will not be repeated here.
The NLL Monte Carlo calculation
of the signal $pp\rightarrow H$ proceeds in a similar manner.

The leading-log signal consists of the Born process ($pp\rightarrow H$)
followed by the decay $H\rightarrow \gamma\gamma$. The Higgs decay
is calculated in the Higgs center-of-mass frame and then boosted into
the hadron-hadron center-of-mass frame.
The Higgs branching fractions are calculated as per
Ref.~\cite{HHG}. Including the
order-$\alpha_{s}$ corrections to the Born process we obtain the NLL signal.
The matrix element is calculated in the approximation of a top quark
with infinite mass. The ``K-factor''
NLO/Born in this limit
is then multiplied by the Born term for a finite top quark mass.
This procedure yields an excellent approximation
to the general case, even for top quark masses above threshold \cite{HQCD3}.

The two photon background consists of several contributions:
Born, gluon box, single- and double-photon fragmentation processes,
and the order-$\alpha_{s}$
corrections to the Born process. The gluon box is an order
$\alpha^{2}\alpha_{s}^{2}$  process but due to
the large gluon luminosity it can not be neglected.
For the remainder of this paper the
leading-log background contribution will be defined as LL = Born + box
+ single fragmentation +
double fragmentation and NLL = LL + order-$\alpha_{s}$ corrections
to the Born process.

Unless otherwise stated the following are used for this calculation:
CTEQ1M parton distributions~\cite{CTEQ1M}, the two-loop expression for
$\alpha_{s}(Q^{2})$, $Q^{2}=p^{2}_{T_{\gamma}}$, and $m_{t}=$ 140 GeV.
Additionally, for the SSC, the following cuts utilized in studies by the
GEM and SDC collaborations~\cite{GEMTDR}
are used: $p_{T_{\gamma}}>$  20 GeV, $|y_{\gamma}|<$ 2.5,
$|\cos\theta^{*}|<$ 0.7, isolation
cone $R\equiv\sqrt{\Delta y^{2}+\Delta\phi^{2}}=0.7$,
and hadronic energy inside of isolation cone $<$ 4 GeV.
For the LHC, the following ATLAS~\cite{ATLAS} inspired cuts are used:
$p^{1}_{T_{\gamma}}>$  30 GeV, $p^{2}_{T_{\gamma}}>$  20 GeV,
$|y_{\gamma}|<$ 2.5,
isolation cone $R\equiv\sqrt{\Delta y^{2}+\Delta\phi^{2}}=0.255$,
$p^{1}_{T_{\gamma}}/(p^{1}_{T_{\gamma}}+p^{2}_{T_{\gamma}})<0.7$
and hadronic energy inside of isolation cone $<$ 5 GeV.

\section{Results}
Defining a $K$-factor as $K =$ NLL/LL, Fig.~1 shows the variation of this
correction factor with photon-pair mass (i.e. Higgs mass) for the SSC
(using GEM/SDC cuts).
The solid curve denotes the variation for the signal and the dashed curve for
the background.
Specific values for the signal and the background may be found in Tables 1-2.

Fig.~1 shows that the situation for a light Higgs (80 GeV $< m_{H} <$ 100 GeV)
may be better
than previously assumed. In this region the $K$-factor for the background
is decreasing while
the $K$-factor for the signal is increasing for decreasing Higgs mass.
The effect of this behavior on the significance
of the signal can be seen in Fig.~2. Fig.~2 shows the ratio of
QCD corrected significance to
the leading-log significance. For the light mass region this curve implies
that the discovery
time may be reduced by up to a factor of 1.4,
and for the rest of the mass region by a factor $\sim$ 1.3.

For the LHC, Fig.~3 shows the variation of the $K$-factor with photon-pair mass
(using ATLAS style cuts).
As before, the solid curve denotes the variation for
the signal and the dashed curve for
the background.
Specific values for the signal and the background may be found in Tables 3-4.

Fig.~3 shows that
the signal $K$-factor dominates the background $K$-factor over the
entire mass range.
The impact of this behavior on the significance
of the signal can be seen in Fig.~4. Fig.~4 shows the ratio
of QCD corrected significance to
the leading-log significance. For the light mass region this curve
implies that the discovery
time may be reduced by up to a factor of 1.8, and for the rest of the
mass region by a factor $\sim$ 1.5.

\section{Conclusions}
Results have been presented, at the NLL level, for
the signal and background in the
intermediate mass region. The $K$-factors, and the
significance of the signal, were found
to depend on the mass of the Higgs boson and the cuts
implemented. The QCD corrections
were found to imply that the discovery time for the
intermediate mass Higgs boson could
be reduced by a factor of 1.3 to 1.8.

%
%
\acknowledgements
This research was supported in part by the U.~S. Department of Energy under
contract numbers DE-FG05-87ER40319 and DE-AC03-76SF00098.
The authors would like to thank I.~Hinchliffe, J.~Womersley and R.~Zhu for
useful discussions.
D.G. acknowledges support by the Max Kade Foundation, New York.
\newpage
%
%
%
%

%
\end{narrowtext}

\newpage
%
%
\figure{$K$-factor for the signal and background at $\sqrt{s}=$ 40 TeV
using GEM/SDC cuts.}
\label{FIG1}
\figure{Ratio NLL significance to LL significance at $\sqrt{s}=$ 40 TeV
using GEM/SDC cuts.}
\label{FIG2}
\figure{$K$-factor for the signal and background at $\sqrt{s}=$ 14 TeV
using ATLAS type cuts.}
\label{FIG3}
\figure{Ratio NLL significance to LL significance at $\sqrt{s}=$ 14 TeV
using ATLAS type cuts.}
\label{FIG4}
%
%
%
%
\widetext
\begin{table}
\caption{Signal (fb) at $\sqrt{s}=$ 40 TeV with GEM/SDC cuts.}
\begin{tabular}{cccc}
$m_{H}$
& $\sigma_{LL} (pp\rightarrow H\rightarrow\gamma\gamma)$
& $\sigma_{NLL} (pp\rightarrow H\rightarrow\gamma\gamma)$
& $K$-factor \\
\tableline
80  & 39 &   67 & 1.72 \\
90  & 49 &   82 & 1.67 \\
100 & 59 &  100 & 1.69 \\
110 & 70 &  117 & 1.67 \\
120 & 76 &  125 & 1.64 \\
130 & 72 &  117 & 1.63 \\
140 & 57 &   93 & 1.63 \\
150 & 37 &   60 & 1.62 \\
160 &  7 &   12 & 1.71 \\
\end{tabular}
\label{TABLE1}
\end{table}
\begin{table}
\caption{Background (fb/GeV) at $\sqrt{s}=$ 40 TeV with GEM/SDC cuts.}
\begin{tabular}{cccc}
$m_{H}$
& $\frac{d\sigma_{LL}(pp\rightarrow\gamma\gamma)}{dM_{\gamma\gamma}}$
& $\frac{d\sigma_{NLL}(pp\rightarrow\gamma\gamma)}{dM_{\gamma\gamma}}$
& $K$-factor \\
\tableline
80  & 761 & 1121 & 1.47 \\
90  & 496 &  780 & 1.57 \\
100 & 340 &  522 & 1.54 \\
110 & 237 &  377 & 1.59 \\
120 & 173 &  283 & 1.64 \\
130 & 128 &  212 & 1.66 \\
140 &  98 &  163 & 1.66 \\
150 &  75 &  127 & 1.69 \\
160 &  60 &  103 & 1.72 \\
\end{tabular}
\label{TABLE2}
\end{table}
\widetext
\begin{table}
\caption{Signal (fb) at $\sqrt{s}=$ 14 TeV with ATLAS type cuts.}
\begin{tabular}{cccc}
$m_{H}$
& $\sigma_{LL} (pp\rightarrow H\rightarrow\gamma\gamma)$
& $\sigma_{NLL} (pp\rightarrow H\rightarrow\gamma\gamma)$
& $K$-factor \\
\tableline
80  & 13.7 & 35.3 & 2.58 \\
90  & 18.3 & 44.5 & 2.43 \\
100 & 22.9 & 53.5 & 2.34 \\
110 & 27.1 & 61.2 & 2.26 \\
120 & 29.2 & 64.9 & 2.22 \\
130 & 27.4 & 59.4 & 2.17 \\
140 & 21.6 & 46.1 & 2.13 \\
150 & 13.8 & 29.1 & 2.11 \\
160 & 2.62 & 5.43 & 2.07 \\
\end{tabular}
\label{TABLE3}
\end{table}
\begin{table}
\caption{Background (fb/GeV) at $\sqrt{s}=$ 14 TeV with ATLAS type cuts.}
\begin{tabular}{cccc}
$m_{H}$
& $\frac{d\sigma_{LL}(pp\rightarrow\gamma\gamma)}{dM_{\gamma\gamma}}$
& $\frac{d\sigma_{NLL}(pp\rightarrow\gamma\gamma)}{dM_{\gamma\gamma}}$
& $K$-factor \\
\tableline
80  & 328 & 667 & 2.03 \\
90  & 249 & 501 & 2.01 \\
100 & 188 & 372 & 1.98 \\
110 & 142 & 286 & 2.01 \\
120 & 108 & 210 & 1.94 \\
130 &  84 & 162 & 1.93 \\
140 &  66 & 127 & 1.92 \\
150 &  53 & 101 & 1.91 \\
160 &  42 &  79 & 1.88 \\
\end{tabular}
\label{TABLE4}
\end{table}
%
%
%
%
\end{document}